%% file: oghpc.tex
\documentclass[9pt]{sig-alternate-05-2015}
\usepackage{listings}
\usepackage[table]{xcolor}
\usepackage{multirow}
\usepackage[american]{babel}
\usepackage[a4paper,margin=0.5in,footskip=0.25in]{geometry}

\selectcolormodel{gray}

\definecolor{dkgreen}{rgb}{0,0.6,0}
\definecolor{gray}{rgb}{0.5,0.5,0.5}
\definecolor{mauve}{rgb}{0.58,0,0.82}
\definecolor{g}{rgb}{1,0.5,0}

\makeatletter
\def\@copyrightspace{\relax}
\makeatother

\begin{document}


\setlength{\pdfpageheight}{\paperheight}
\setlength{\pdfpagewidth}{\paperwidth}




\title{A Survey of Sparse Matrix-Vector Multiplication Performance on Large Matrices}

\numberofauthors{3}

\author{
    \alignauthor
    Max Grossman\\
      \affaddr{Rice University}\\
      \email{jmg3@rice.edu}
    \alignauthor
    Christopher Thiele\\
      \affaddr{Shell International Exploration \& Production Inc.}\\
    \alignauthor
    Mauricio Araya-Polo\\
      \affaddr{Shell International Exploration \& Production Inc.}\\
    \and \\
    \alignauthor
    Florian Frank\\
      \affaddr{Rice University}\\
    \alignauthor
    Faruk O. Alpak\\
      \affaddr{Shell International Exploration \& Production Inc.}\\
    \alignauthor
    Vivek Sarkar\\
      \affaddr{Rice University}\\
}

\maketitle




\section{Motivation}
\label{sec:motivation}
\input{motivation}

\section{Background}
\label{sec:background}
\input{background}

\section{Methods}
\label{sec:methods}
\input{methods}

\section{Experimental Evaluation}
\label{sec:eval}
\input{eval}

\section{Conclusions}
\label{sec:concl}
\input{concl}


\bibliographystyle{abbrv}
\bibliography{cited}  

\end{document}

%% file: motivation.tex
One of the main sources of sparse matrices is the discretization of partial differential equations that govern continuum-physics phenomena such as fluid flow and transport, phase separation, mechanical deformation, electromagnetic wave propagation, and others.  Recent advances in high-performance computing area have been enabling researchers to tackle increasingly larger problems leading to sparse linear systems with hundreds of millions to a~few tens of billions of unknowns, e.g., \cite{dogru2009next,Dogru2011NewFrontiers}.  Iterative linear solvers are popular in large-scale computing as they consume less memory than direct solvers.  
Contrary to direct linear solvers, iterative solvers approach the solution gradually requiring the computation of sparse matrix-vector (SpMV) products.
The evaluation of SpMV products can emerge as a~bottleneck for computational performance within the context of the simulation of large problems. 
In this work, we focus on a~linear system arising from the discretization of the Cahn--Hilliard equation, which is a~fourth order nonlinear parabolic partial differential equation that governs the separation of a~two-component mixture into phases~\cite{CahnHilliard1958}. The underlying spatial discretization is performed using the discontinuous Galerkin method~(e.g.~\cite{riviere2008}) and Newton's method. 
A~number of parallel algorithms and strategies have been evaluated in this work to accelerate the evaluation of SpMV products.

%% file: background.tex
There exist many factors that significantly impact the performance of a~given SpMV computation. We separate these factors into four categories: matrix
characteristics, storage format, software implementation, and hardware platform.

Matrix characteristics of the input matrix include:
its definiteness, eigenvalues, number of non-zero values (NNZ), and imbalances in terms of NNZ.
While some matrix characteristics are immutable
at runtime others can be changed to improve
performance. For example, sorting the rows of a sparse matrix by their NNZ can
reduce NNZ imbalance between neighboring rows.

There are many ways to store a~sparse matrix, including Compressed Sparse
Row (CSR), Compressed Sparse Column (CSC), Coordinate (COO), Diagonal (DIA),
Ellpack-Itpack\\ (ELL), Sliced ELL (SELL), Hybrid (HYB), Blocked CSR (BSR), and
Extended BSR (BSRX). Each of these formats differs in its space consumption,
lookup efficiency, redundant zero values, and implementation complexity.
Additionally, the best storage format for a given matrix is often dependent on
characteristics of the matrix itself. The matrix format can almost always be freely
tuned to optimize space or time efficiency.

There are several optimized SpMV implementations available. In our
work, we focus on a preliminary evaluation of Intel MKL~\cite{mkl}, the Trilinos
project~\cite{trilinos}, CUSPARSE~\cite{cusparse}, and CUSP~\cite{cusp}.
While application developers are generally free to choose whichever SpMV
implementation they prefer (ignoring licensing constraints), that choice
generally places hard constraints on the problems that can be solved. For
example, libraries may only support SpMV on a limited number of storage formats,
may only be available or optimized on a certain hardware platform, or may limit
the size of the stored matrix by their choice of type used to represent matrix
size.

Some SpMV implementations are closely tied to the hardware they are executed on (e.g.~MKL, CUSPARSE). Therefore, the choice of hardware platform generally constrains
the choice of software implementation, and vice versa. The
characteristics of a hardware platform (e.g.~DRAM bandwidth, cache hierarchy,
parallelism) can have a significant impact on SpMV performance. Past work
has generally found that SpMV is a memory-bound kernel whose performance is heavily
tied to the cache hierarchy and memory bandwidth of the hardware platform it
runs on~\cite{williams2009optimization}.

In this work, we contribute a third-party survey of SpMV performance on
industrial-strength, large matrices using:

\begin{enumerate}
\item The SpMV implementations in Intel MKL, the Trilinos project (Tpetra
      subpackage), the CUSPARSE library, and the CUSP library, each running on
      modern architectures.
\item NVIDIA GPUs and Intel multi-core CPUs (supported by each software
      package).
\item The CSR, BSR, COO, HYB, and ELL matrix formats (supported by each
      software package).
\end{enumerate}

%% file: methods.tex
\subsection{Matrix characterization}
The origin of the matrices used in this work is described in Section~\ref{sec:motivation}. 
We use six different not diagonal (but band) matrices of varying sizes and characteristics
in this work. Selected matrix characteristics are listed in
Table~\ref{tab:matrices}. Each matrix has a mode of 10 non-zeroes per row, and a
maximum of 16 non-zeroes per row.


\begin{table}
\centering
\small
\begin{tabular}{ | c | c | c | c | c | }
\hline
\textbf{Matrix} & \textbf{Dim} & \textbf{\% NZ} & \textbf{Avg NZ/row} & \textbf{CSR Size} \\\hline
A & 611K    & 0.00172\% & 10.53 & 79.74 MB  \\\hline
B & 1,555K  & 0.00068\% & 10.48 & 201.83 MB \\\hline
C & 4,884K  & 0.00023\% & 11.01 & 664.73 MB \\\hline
D & 12,434K & 0.00009\% & 10.98 & 1.689 GB  \\\hline
E & 39,079K & 0.00003\% & 11.25 & 5.434 GB  \\\hline
F & 99,476K & 0.00001\% & 11.24 & 13.818 GB \\\hline
\end{tabular}
\caption{Intrinsic characteristics for each matrix tested.}
\label{tab:matrices}
\end{table}

\newpage
\subsection{Experimental platform}
All experimental evaluations are performed on a hardware platform containing
a 12-core 2.80GHz Intel X5660 CPU with 48GB of system RAM paired with an NVIDIA
M2070 with $\sim$2.5GB of graphics memory. The configurations used for each
software platform are listed in Table~\ref{tab:software}.

\begin{table}
\centering
\begin{tabular}{ | c | c | c | c | }
\hline
\textbf{Software} & \textbf{Version} & \textbf{Proc.} & \textbf{Matrix Formats} \\\hline
MKL      & 11.2.2.164 & CPU & CSR, COO, BSR \\\hline
Trilinos & 12.2.1     & CPU & CSR \\\hline
CUSPARSE & 6.5        & GPU & CSR, BSR, HYB, ELL \\\hline
CUSP     & 0.5.1      & GPU & CSR, COO, HYB, ELL \\\hline
\end{tabular}
\caption{Configuration of each software platform.}
\label{tab:software}
\end{table}




%% file: eval.tex
Here we present preliminary results evaluating the relative performance of MKL,
Trilinos, CUSPARSE, and CUSP.

We start our evaluation by identifying the optimal matrix format for each software package,
with varying numbers of threads for the CPU-based packages. For each
combination of software package and matrix we perform 30 runs and select the fastest format based on the
median. For BSR, we try all blocking factors less than or equal to 32 that
evenly divide the matrix dimensions. For HYB, we try with the ELL portion's
width set 10, 11, 12, 13, 14, and 15.
Some formats are more space efficient than
others, and so not all formats could be evaluated on all matrices due to
out-of-memory errors.

We find that the most efficient format for all matrices and all thread counts
running on MKL is CSR. Trilinos only supports the CSR format, so no matrix
format comparison is necessary. For CUSPARSE and CUSP the optimal matrix format
for all matrices is also CSR. This result may be surprising as past work 
highlights the benefits of the ELL and HYB formats for
GPUs~\cite{bell2009implementing}. However, the small
number of non-zeroes per row in our test matrices reduces memory coalescing in
vectorized formats like ELL and HYB.


Next, we compare the best performance achievable on each software platform for
each matrix. Table~\ref{tab:execution-speedup} lists the maximum speedup each
platform achieves, relative to MKL single-threaded. These speedup numbers
including copying the inputs and outputs to and from the GPU for CUSPARSE and
CUSP. Note the slowdown of both GPU frameworks, relative to single-threaded MKL,
which contrasts with most or all of the related
literature~\cite{bell2009implementing,davis2012spmv}. To explain this, we
compare kernel execution times in Table~\ref{tab:kernel-speedup} (ignoring GPU
communication). There, we see results more similar to past evaluations, with
CUSPARSE achieving $\sim$10$\times$ speedup relative to MKL single-threaded and
an average of 3.123$\times$ speedup relative to the fastest CPU implementation
for each matrix.

\begin{table}
\centering
\begin{tabular}{ | c | c | c | c | c | }
\hline
\textbf{Matrix} & \textbf{MKL} & \textbf{Trilinos} & \textbf{CUSPARSE} & \textbf{CUSP} \\\hline
A & 3.87 (8T) & 3.65 (8T) & 0.89 & 0.59 \\\hline
B & 2.97 (8T) & 3.16 (8T) & 0.83 & 0.60 \\\hline
C & 2.84 (8T) & 3.16 (8T) & 0.79 & 0.58 \\\hline
D & 3.15 (8T) & 2.54 (2T) & 0.87 & N/A \\\hline
E & 3.33 (8T) & 4.69 (8T) & 0.84 & N/A \\\hline
F & 3.52 (8T) & 4.75 (2T) & 0.82 & N/A \\\hline
\end{tabular}
\caption{Speedups for each software platform on each matrix, relative to
single-threaded MKL, where NT~indicates that the best performance was achieved with
N~CPU threads. Missing data for CUSP is due to the difficulty of
supporting out-of-core datasets in its API.}
\label{tab:execution-speedup}
\end{table}


\begin{table}
\centering
\begin{tabular}{ | c | c | c | }
\hline
\textbf{Matrix} & \textbf{CUSPARSE} & \textbf{CUSP} \\\hline
A & 11.20 & 4.11 \\\hline
B & 10.41 & 3.70 \\\hline
C & 10.47 & 3.76 \\\hline
D & 10.35 & N/A  \\\hline
E & 10.76 & N/A  \\\hline
F & 10.81 & N/A  \\\hline
\end{tabular}
\caption{Speedups for CUSPARSE and CUSP relative to single-threaded MKL, kernel time only.}
\label{tab:kernel-speedup}
\end{table}

%% file: concl.tex
The evaluation performed in this work is significant for a number of reasons:

\begin{enumerate}
\item Even though MKL is 
      widely use for sparse linear algebra and
      offers a low-level C API, Trilinos is able to match or beat MKL
      performance on several of our matrices, particularly larger ones.
      Therefore, using Trilinos's flexible, object-oriented API
      becomes the preferred choice without having to worry about sacrificing performance.
\item When only considering kernel performance, CUSPARSE is able to demonstrate 3.123$\times$ speedup relative to the best
      CPU-based packages. However, these benefits immediately disappear when
      data movement is considered. 
Emphasis must be placed on
      keeping data on the devices or maximizing computation-communication
      overlap. As CUSPARSE gives more control over communication than CUSP, it
      is the preferred GPU framework.
\end{enumerate}

These preliminary results make it clear that we should focus our future
investigation and work around Trilinos and CUSPARSE. Our next steps will focus
on evaluating Trilinos's GPU support, extending this work to distributed
systems, considering multi-GPU execution, experimenting with the MKL
inspector-executor framework, and investigating Trilinos-CUSPARSE integration.

%% file: oghpc.bbl
\begin{thebibliography}{10}

\bibitem{cusp}
{CUSP}.
\newblock \url{https://github.com/cusplibrary/cusplibrary}.

\bibitem{bell2009implementing}
N.~Bell and M.~Garland.
\newblock Implementing sparse matrix-vector multiplication on
  throughput-oriented processors.
\newblock In {\em Proceedings of the Conference on High Performance Computing
  Networking, Storage and Analysis}, page~18. ACM, 2009.

\bibitem{CahnHilliard1958}
J.~W. Cahn and J.~E. Hilliard.
\newblock Free energy of a nonuniform system. {I.} {I}nterfacial free energy.
\newblock {\em The Journal of Chemical Physics}, 28:258--267, 1958.

\bibitem{davis2012spmv}
J.~D. Davis and E.~S. Chung.
\newblock Spmv: A memory-bound application on the gpu stuck between a rock and
  a hard place.
\newblock {\em Microsoft Research Silicon Valley, Technical Report14
  September}, 2012, 2012.

\bibitem{dogru2009next}
A.~H. Dogru, L.~S. Fung, U.~Middya, T.~M. Al-Shaalan, J.~A. Pita,
  K.~HemanthKumar, H.~Su, J.~C. Tan, H.~Hoy, W.~Dreiman, et~al.
\newblock A next-generation parallel reservoir simulator for giant reservoirs.
\newblock In {\em SPE/EAGE Reservoir Characterization \& Simulation
  Conference}, 2009.

\bibitem{Dogru2011NewFrontiers}
A.~H. Dogru, L.~S.~K. Fung, U.~Middya, T.~M. Al-Shaalan, T.~Byer, H.~Hoy, W.~A.
  Hahn, N.~Al-Zamel, J.~A. Pita, K.~Hemanthkumar, M.~Mezghani, A.~Al-Mana,
  J.~Tan, W.~Dreiman, A.~Fugl, and A.~Al-Baiz.
\newblock New frontiers in large scale reservoir simulation.
\newblock {\em Society of Petroleum Engineers}, 2011.

\bibitem{mkl}
Intel.
\newblock {Intel Math Kernel Library}.
\newblock \url{https://software.intel.com/en-us/intel-mkl}.

\bibitem{trilinos}
S.~N. Laboratories.
\newblock {The Trilinos Project}.
\newblock \url{https://trilinos.org/}.

\bibitem{cusparse}
NVIDIA.
\newblock {CUSPARSE}.
\newblock \url{https://developer.nvidia.com/cusparse}.

\bibitem{riviere2008}
B.~Rivi\`ere.
\newblock {\em Discontinuous Galerkin Methods for Solving Elliptic and
  Parabolic Equations: Theory and Implementation}.
\newblock Frontiers in Applied Mathematics. Society for Industrial and Applied
  Mathematics, 2008.

\bibitem{williams2009optimization}
S.~Williams, L.~Oliker, R.~Vuduc, J.~Shalf, K.~Yelick, and J.~Demmel.
\newblock Optimization of sparse matrix--vector multiplication on emerging
  multicore platforms.
\newblock {\em Parallel Computing}, 35(3):178--194, 2009.

\end{thebibliography}
